\documentclass[aps,prl,twocolumn,showpacs,floats]{revtex4}
\usepackage[dvips]{graphicx}
\usepackage{amsmath}
\newcommand{\be}{\begin{equation}}
\newcommand{\ee}{\end{equation}}
\newcommand{\bea}{\begin{eqnarray}}
\newcommand{\eea}{\end{eqnarray}}
\newcommand{\ba}{\begin{array}}
\newcommand{\ea}{\end{array}}
\newcommand{\nn}{\nonumber}

\newcommand{\Del}{\Delta}
\newcommand{\del}{\delta}

\newcommand{\gam}{\gamma}
\newcommand{\al}{\alpha}
\newcommand{\sig}{\sigma}

\newcommand{\noi}{\noindent}
\newcommand{\eps}{\epsilon}
\newcommand{\lam}{\lambda}
\newcommand{\ua}{\uparrow}
\newcommand{\da}{\downarrow}

\newcommand{\ra}{\rangle}
\newcommand{\la}{\langle}
\begin{document}

\title{\bf Implementation of the quantum walk step operator in lateral quantum dots}

\author{K.A. van Hoogdalem$^{1,2}$ and M. Blaauboer$^{1}$}

\affiliation{$^{1}$ Kavli Institute of Nanoscience, Delft University of Technology,
Lorentzweg 1, 2628 CJ Delft, The Netherlands}
\affiliation{$^{2}$Department   of  Physics,   University   of  Basel, Klingelbergstrasse 82, CH-4056 Basel, Switzerland}

\date{\today}

\begin{abstract}
We propose a physical implementation of the step operator of the discrete quantum 
walk for an electron in a one-dimensional chain of quantum dots. The operating 
principle of the step operator is based on locally enhanced Zeeman splitting and
the role of the quantum coin is played by the spin of the electron. 
We calculate the probability of successful transfer of the electron in 
the presence of decoherence due to quantum charge fluctuations, modeled as a 
bosonic bath. We then analyze two mechanisms for creating locally 
enhanced Zeeman splitting based on, respectively, locally applied electric and magnetic 
fields and slanting magnetic fields. Our results imply that 
a success probability of $>$ 90\% is feasible under realistic experimental conditions.
\end{abstract}

\maketitle

The quantum walk (or quantum random walk) is the quantum-mechanical analogue of the 
classical random walk and describes the random walk behavior of a quantum particle. 
The concept "quantum walk" was formally introduced by Aharonov {\it et al.} in 
1993~\cite{ahar93} and suggested earlier by Feynman~\cite{feyn65}. The essential 
difference with the classical random walk lies in the role of the coin: Whereas in the 
classical random walk the coin is a classical object with two possible measurement 
outcomes ("heads" or "tails"), in the quantum walk the coin is a quantum-mechanical 
object - typically a two-level system such as a 
spin-1/2 particle - which can be measured along different bases and hence has a 
multi-sided character. As a result, the quantum walk exhibits strikingly different 
dynamic behavior compared to its classical counterpart due to interference between 
different possible paths. One example is faster propagation~\cite{ahar93}: The 
root-mean-square distance from the origin $\la x\ra_{\rm rms}$ that is covered by 
a quantum walker grows linearly 
with the number of steps $N$ (thus corresponding to ballistic propagation), 
whereas for the classical random walk 
$\la x \ra_{\rm rms} \sim \sqrt{N}$ (corresponding to diffusive propagation). This property 
has been exploited to design new quantum computing algorithms~\cite{kend06}. 
Both discrete and continuous time quantum walks have been extensively studied 
in recent years~\cite{kemp03}, including investigations of decoherence~\cite{brun03} 
and entanglement between quantum walkers~\cite{vene05}.

As far as implementations of quantum walks in actual physical systems are concerned, 
several proposals have been put forward for a range of optical and atomic 
systems, such as optical cavities~\cite{knig03}, cavity QED systems~\cite{sand03}, 
trapped atoms and ions~\cite{trav02} and linear optical elements~\cite{path07}. 
On the experimental side, only a few realizations of quantum walks have been 
achieved: Discrete and continous quantum walks in NMR quantum systems~\cite{du03}, 
discrete quantum walks using linear optical elements~\cite{do05} and, most 
recently, a continuous quantum walk in an optical waveguide lattice~\cite{pere07}. 

For solid-state systems, no realizations of quantum walks exist so far. A recent proposal 
for implementation of a quantum algorithm using NAND operations in a tree of quantum dots 
relies on the continuous time quantum walk~\cite{tayl07}. 

In this paper we propose the first implementation of a discrete quantum walk in a 
solid-state quantum system, which consists of a single electron traveling in a one-dimensional 
chain of quantum dots. In particular, we focus on the implementation of the so-called step operator, 
the basic unit of the quantum walk. The step operator causes the electron to either move to 
the left or to the right depending on the state of the quantum coin, 
which in our model is represented by the spin of the electron. We calculate the spin-dependent 
transfer probability 
of the electron from one dot to the next in the presence of different energy level splittings 
in neighboring quantum dots (due to locally enhanced Zeeman splitting), taking into account 
the effects of decoherence due to gate voltage fluctuations~\cite{dephasing}. 
We then propose two physical mechanisms 
to achieve locally enhanced level splitting in a quantum dot using local electric 
and magnetic fields and find that under current experimental circumstances successful implementation
of the step operator is possible with $>$ 90\% probability.

\begin{figure}
\centering
\includegraphics[width=0.9\columnwidth]{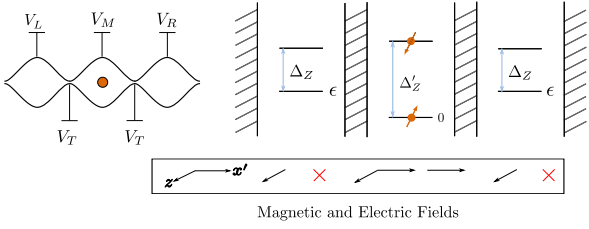}
\caption[]{(left) A linear chain of 3 quantum dots in a magnetic field 
and (right) the corresponding energy level diagram. The direction
of (locally) applied electric and magnetic fields in each dot is also
indicated, a cross indicates no applied field.
See the text for further explanation.} 
\label{fig:model}
\end{figure}

{\it Model of the step operator. -} Consider a chain of three quantum dots in series, 
in which the middle dot (M) is occupied by a single electron, see Fig.~\ref{fig:model}.
The energy level diagram shows the Zeeman splitting $\Del_z^{(\prime)}$ of the 
lowest two levels in a magnetic field, and we assume that
this splitting is larger in the middle dot by an amount $\Del_z^{\prime} - \Del_z$. The gate voltages 
$V_L$, $V_M$ and $V_R$ are used to shift the energy levels in the left, middle and right dot 
resp., and the gate voltages $V_T$ are used to tune the tunnel coupling between 
neighboring dots. The initial spin state of the electron is a superposition of spin-$\uparrow$ 
and spin-$\downarrow$ and our first goal is to design an implementation of the step operator 
such that it causes the electron to coherently move to the left (right)
if it has spin-$\uparrow$ (spin-$\downarrow$) and to calculate the probability of succesful 
transfer for this process. To begin with, we assume the right dot to be decoupled 
from the other two and calculate the spin-dependent probabilities to find the electron in the 
left and middle dot at a given time. In the absence of decoherence - which is considered further 
below - the Hamiltonian for the spin-$\uparrow$ and spin-$\downarrow$ components of the electron 
is given by
\be
H_{0,\sigma} = \left( \ba{cc}
A_{\sigma} & \lam \\
\lam & B_{\sigma}
\ea \right),
\label{eq:Ham}
\ee
in the basis $\{ |L\ra, |M\ra \}$. Here ($A_{\ua}$,$B_{\ua}$) = ($\eps$,0), 
($A_{\da}$,$B_{\da}$) = ($\eps + \Del_z$,$\Del_z^{'}$) and $\lam$ is the (spin-independent) tunnel coupling between the left and middle dot.
The eigenvalues and -vectors of Eq.~(\ref{eq:Ham}) are given by $E_{\pm, \sig} = \frac{A_{\sig}+B_{\sig}}{2} \pm \frac{1}{2}
\sqrt{(A_{\sig}-B_{\sig})^2 + 4 \lam^2}$, $|\psi_{+,\sig}\ra = (\sin \theta_{\sig}, \cos \theta_{\sig})^{T}$ and  
$|\psi_{-,\sig}\ra = (\cos \theta_{\sig}, -\sin \theta_{\sig})^{T}$, with $\tan \theta_{\sig} \equiv 
\frac{\hbar \del_{\sig} + \sqrt{(\hbar \del_{\sig})^2 + 4\lam^2}}
{2\lam}$, $\del_{\ua} \equiv \eps/\hbar$ and $\del_{\da} = (\eps + \Del_z - \Del_z^{'})/\hbar$.
In the absence of decoherence, the solution of the density matrix equations  
$\dot{\rho} = -(i/\hbar) [H_{0,\sig},\rho]$ for the population in the left dot is given by
\be
\rho_{LL,\sig}(t) = \frac{2 \lam^2}{(\hbar \omega_{\sig})^2} \left[ 1 - \cos (\omega_{\sig} t) \right]
\label{eq:solution}
\ee
for initial conditions $\rho_{LL,\sig} (0)=0$, $\rho_{MM,\sig} (0)=1$ and $\hbar \omega_{\sig} \equiv 
\sqrt{(\hbar \del_{\sig})^2 + 4\lam^2}$. 
We see that the probability for the spin-$\ua$ component to be in the left 
dot and the spin-$\da$ component to remain in the middle dot is 1 for $\frac{(\hbar \omega_{\ua})^2}{2 \lam^2} = 
1 - \cos \left( 2n \pi \frac{\omega_{\ua}}{\omega_{\da}} \right)$, $n=0,1,2,\ldots$. The smallest $n$ that
yields a solution to this equation for $\eps=0$ (zero detuning) is $n=1$, for which $\Del_z - \Del_z^{\prime} = 
2\sqrt{3}\, \lam$ and 
$t=2\pi/\omega_{\da}$. The second half of the step operator then consists of repeating the same 
procedure as described above 
for tunnel coupling between the middle and the right dot, now assuming the left dot to be decoupled.

\begin{figure}
\centering
\includegraphics[width=0.98\columnwidth]{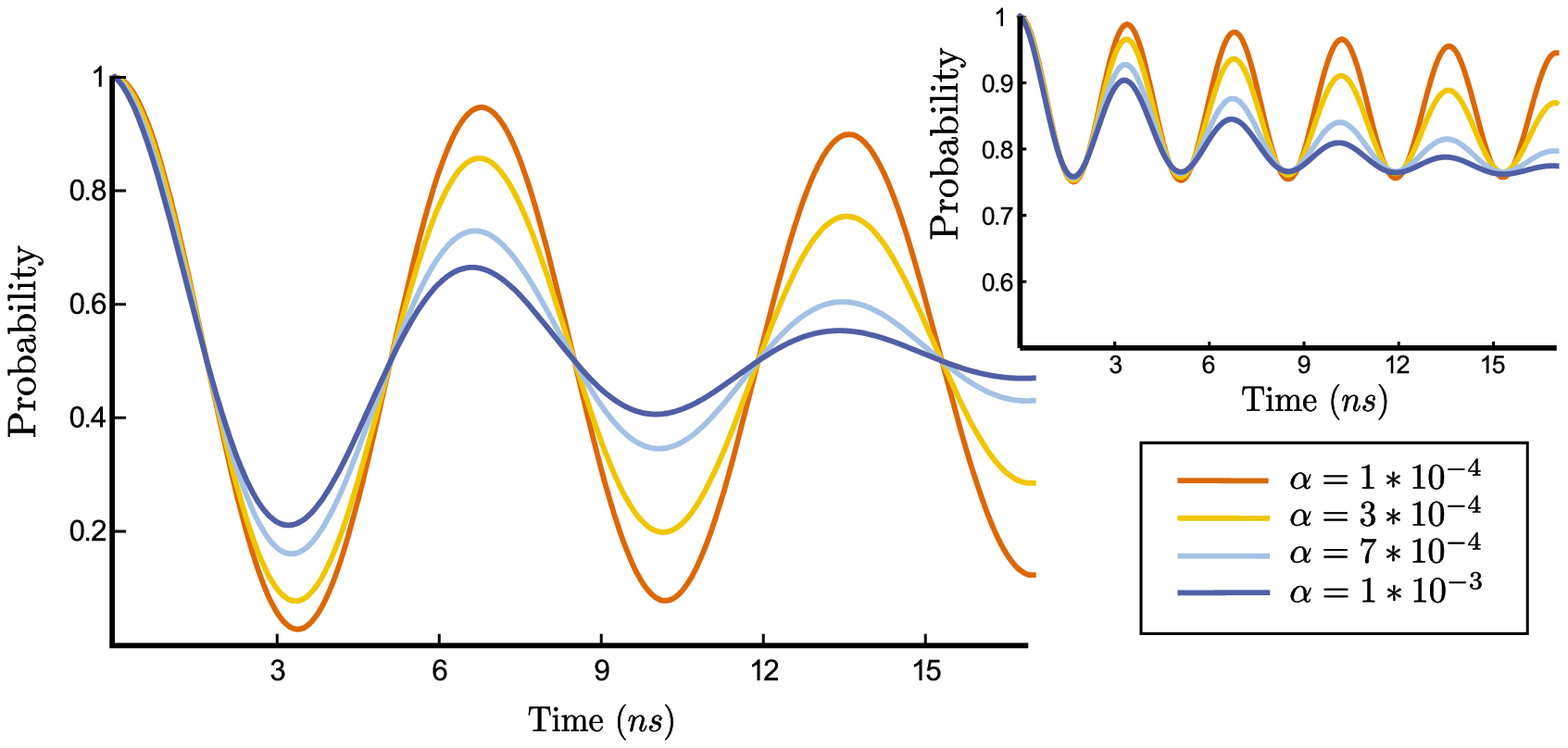}
\caption[]{Survival probability [Eq.~(\ref{eq:survival})] of the spin-$\uparrow$-component (main plot)
and spin-$\downarrow$-component (inset) in the middle dot as a function of time $t$ for
various values of $\al \equiv \al_{\lam} = \al_{\eps}$~\cite{asta04}. Parameters 
used are $\hbar \del_{\ua} = 0$, $\hbar \del_{\da}  = 1 \mu$eV, 
$\lam= \left(2\sqrt{3}\right)^{-1}\, \mu$eV~\cite{kopp05}, and $T = 10$ mK.
}
\label{fig:survival}
\end{figure}

{\it Decoherence due to quantum fluctuations. -} In reality, the time evolution of the occupation probabilities 
$\rho_{\sigma}$ is affected by decoherence due to coupling of the quantum dots to the 
environment. In typical experimental situations $k_B T < \hbar \omega_{\sig}$, for $T \leq$ 100 mK and 
$\lam \leq 1.5\, 10^{-24}$ J~\cite{pett05,kopp05}, 
so that quantum noise, rather than classical noise, is the dominant source of 
decoherence. Specifically, since the quantum walk step operator based on the Hamiltonian (\ref{eq:Ham})
involves tuning the tunnel couplings $\lam$ - resulting in different charge occupations on neighboring dots - 
the probability distributions will be strongly affected by fluctuations of charge in the 
environment~\cite{spinfluctuations}. The quantum charge fluctuations we consider here 
are gate voltage fluctuations, which cause both fluctuations in the tunnel coupling 
$\lam$ and in the energy levels $E_{\pm}$ of 
the Hamiltonian (\ref{eq:Ham})~\cite{romi07}. 
The Hamiltonian which describes the quantum dot system plus the environment is given by
\be
H = H_0 + V_{\eps} A_{\eps} +  V_{\lam} A_{\lam} + 
H_{bath,\eps} + H_{bath,\lambda},
\label{eq:fullHam}
\ee
with $H_0$ the Hamiltonian (\ref{eq:Ham}) of the isolated system, $V_{\lam}$ =
$|L\ra \la M| + |M\ra \la L|$, $V_{\eps}$ = $|L\ra \la L| - |M\ra \la M|$, 
$A_{\eps (\lam)}$ = $\sum_{k} c_{k,\eps (\lam)} ( b_{k,\eps (\lam)}^{\dagger} + b_{k,\eps (\lam)})$
and $H_{bath,\eps (\lam)}$ = $\sum_{k} \hbar \omega_{k,\eps (\lam)} 
b_{k,\eps (\lam)}^{\dagger} b_{k,\eps (\lam)}$. We model the environment as a bosonic bath~\cite{weis99}
with creation and annihilation operators $b_{k,\eps (\lam)}^{\dagger}$ and 
$b_{k,\eps (\lam)}$ and use the Born-Markov approximation~\cite{cohe92} to calculate
the time evolution of the spin-dependent occupation probabilities under the
Hamiltonian (\ref{eq:fullHam}), assuming weak coupling between the system and the environment 
and short correlation times of the boson baths (Markovian assumption). The baths are characterized by 
the symmetric and 
anti-symmetric spectral functions $S^{\pm}(\omega)$ with 
$S^+(\omega)=\coth \left( \frac{\hbar \omega}{2 k_B T} \right) S^-(\omega)$~\cite{weis99} and
we assume Ohmic baths with Lorentzian damping for which $S^-(\omega) = \alpha \hbar^2 \omega 
\frac{1}{(\omega/\omega_c)^2+1}$ with $\omega_c$ a cut-off frequency. The time evolution of the 
reduced density matrix $\rho_{ab}(t)$ is then given by $\frac{\mathrm{d} \rho_{ab}(t)}{\mathrm{d} t} 
= -i \omega_{ab} \rho_{ab}(t) 
+ \sum_{cd}^{\textrm{(sec)}} R_{abcd} \rho_{cd} (t)$, with $\omega_{ab} \equiv (E_a - E_b)/\hbar$
and $R_{abcd}$ the Bloch-Redfield tensor~\cite{weis99,limitsum}. Solving this master equation for the two-level
system $\{ |\psi_{+,\sig}\ra, |\psi_{-,\sig}\ra \}$ described above for an electron starting at 
$t=0$ in the middle dot yields for the evolution of the population $\rho_{--}$ of the groundstate $|\psi_{-}\ra$
and the coherence terms $\rho_{-+}$:
\begin{subequations}
\bea
\rho_{--,\sig}(t) & = & \frac{1}{2} ( \tanh ( \frac{\hbar \omega_{\sig}}{2 k_B T}) 
- \cos(2 \theta_{\sig}) ) e^{-\gam_1 t} + \nn \\
& & \frac{\coth ( \frac{\hbar \omega_{\sig}}{2 k_B T}) - 1}{2 \coth ( \frac{\hbar \omega_{\sig}}{2 k_B T})}
\label{eq:sigma_minus} \\
\rho_{-+,\sig}(t)  & = &  \rho_{+-,\sig}^{*}(t) = 
- \sin \theta_{\sig} \cos \theta_{\sig} e^{-i \omega_{\sig} t} e^{-\gam_2 t}
\eea
\label{eq:pop_coh}
\end{subequations}
with
\begin{subequations}
\bea
\gam_1 & = & 4 \pi^2 \left( \alpha_{\lambda} \cos^2 2 \theta_{\sig} + \alpha_{\eps} \sin^2 2 
\theta_{\sig} \right) 
\omega_{\sig} \coth \left( \frac{\hbar \omega_{\sig}}{2 k_B T} \right) \\
\gamma_2 & = & \frac{\gam_1}{2} + \frac{8 \pi^2}{\hbar} \left( \alpha_{\lambda} \sin^2 2 \theta_{\sig} + 
\alpha_{\eps} \cos^2 2 \theta_{\sig} \right) k_B T.
\eea
\label{eq:gammas}
\end{subequations}
\noi Eqns.~(\ref{eq:pop_coh}) and (\ref{eq:gammas}) are 
valid in the limit $\gam_1$, $\gam_2$ $\ll |\hbar \omega_{\sig}|, 1/\bar{\tau}$ 
(with $\bar{\tau}$ the bath correlation time)~\cite{cohe92}. The survival probability to remain
in the middle dot at time $t$ is then given by 
\bea
P_{M,\sig}(t) & = & 
\frac{1}{2} \left[ 1 + \cos (2\theta_{\sig}) \tanh \left( \frac{\hbar \omega_{\sig}}{2 k_B T} \right)
( 1 - e^{-\gam_1 t}) + \right. \nn \\
& & \left. \cos^2 (2\theta_{\sig})\, e^{-\gam_1 t} + 
\sin^2 (2\theta_{\sig}) \cos (\omega_{\sig} t)\, e^{-\gam_2 t}
\right]
\label{eq:survival}
\eea
and plotted for both spin-$\uparrow$ and spin-$\downarrow$ in Fig.~\ref{fig:survival}.
Using typical experimental parameters (see figure caption and discussion below) we see that succesful transfer of 
the spin-$\uparrow$
component to the left dot while spin-$\downarrow$ remains in the middle dot occurs with probability $>$ 90\%
(for $\al \leq 3 \cdot 10^{-4}$) at $t\approx 3.5$ ns.

{\it Locally enhanced Zeeman splitting. -} We now consider the question how a different Zeeman 
splitting in neighboring dots - as assumed in the calculations above - can be
realized in practice. In particular, we propose two mechanisms to achieve locally enhanced Zeeman 
splitting. The first one is by application of a local transverse magnetic field and a local electric field 
and relies on spin-orbit interaction. Both tunable local magnetic and electric fields have recently 
been demonstrated experimentally~\cite{kopp06,nowa07}. In our model, see 
Fig.~\ref{fig:model}, we assume that a local magnetic field (in addition to the global
Zeeman-splitting field $\vec{B}$) and a local electric field $\vec{E}(t)$ are applied to the middle quantum dot.
In the presence of spin-orbit interaction, the Hamiltonian for this middle dot 
is then given by $H = H_0 + e \vec{E}(t) \cdot \vec{r} + \alpha \left( p_x \sigma_y - p_y \sigma_x \right) 
+ \beta \left( -p_x \sigma_x + p_y \sigma_y \right)$, with $H_0$ the Hamiltonian given by  
Eq.~(\ref{eq:Ham}), and $\al$ and $\beta$ resp. the Rashba and Dresselhaus spin-orbit coupling 
strengths. Using the Schrieffer-Wolff transformation, the Hamiltonian $H$ can be diagonalized 
to first order in the spin-orbit terms which 
yields, for the spin-dependent part, $H_{\rm eff} \equiv e^{S}He^{-S} =  \frac{1}{2} g^{*} \mu_B 
\left( \vec{B} + \delta \vec{B}(t) \right) \cdot \vec{\sigma}$~\cite{borh06}. Here the effective magnetic field
$\del \vec{B}$ is given by $\delta \vec{B} = 2 \vec{B} \times \left( \vec{\Omega}_1(t) + \vec{n} 
\times \vec{\Omega}_2(t) \right)$, with $\vec{\Omega}_1(t) = \frac{e \hbar}{E_Z} \alpha_1 
\left( E_{y'}/\lambda_-,E_{x'}/\lambda_+,0\right)$, $\vec{\Omega}_2(t) =  \frac{e \hbar}{E_Z} \beta_1 
\left( -E_{x'}/\lambda_-,E_{y'}/\lambda_+,0\right)$, $\al_1 = \frac{\hbar}{m^{*}} \frac{E_Z \left( E_Z^2 - 
(\hbar \omega_0)^2\right)}{(E_Z^2 - E_1^2)(E_Z^2-E_2^2)}$, $\beta_1  =  \frac{\hbar}{m^{*}}\frac{E_Z^2 
\hbar \omega_c}{(E_Z^2 - E_1^2)(E_Z^2-E_2^2)}$, $\lam_{\pm} = \hbar/(m^{*}(\beta \pm \alpha))$,
$E_Z = g^{*} \mu_B B$, $E_{1,2} = \hbar ( \sqrt{4\omega_0^2 + \omega_c^2} \pm \omega_c)/2$,
with $\omega_c = e B /m^{*}$ the cyclotron frequency and $m^{*} \omega_0^2 r^2/2$ the 
harmonic potential of the quantum dot~\cite{hans07}. In these expressions we have used $x'\equiv (x+y)/\sqrt{2}$ and $y'\equiv (y-x)/\sqrt{2}$. Since for the step operator we require the eigenstates of the Hamiltonian $H$ to be $|\ua\ra$ and $|\da\ra$ we must account for the fact that the eigenvectors of the effective Hamiltonian $H_{\rm eff}$ are transformed by $e^{S}$ with respect to the eigenvectors of $H$ corresponding to the same eigenvalues. We do this be requiring that the eigenvectors of $H_{\rm eff}$ are $e^{S}|\ua\ra$ and $e^{S}|\da\ra$ to first order in the action $S$.

Let us assume the global magnetic field $\vec{B}$ to be parallel to the $\hat{z}$-axis. From the 
expression for $H_{\rm eff}$ it follows that in order to locally generate a different Zeeman 
splitting (along the $\hat{z}$-axis) in the middle dot compared to its neighboring dots, an additional 
tunable local magnetic field, e.g. along the $\hat{x}'$-axis, and a tunable local electric field, also along the $\hat{x}'$-axis, are required. We define the total (local) magnetic field $\vec{B}(t) = B_0\left(n_{x'}(t),0,n_z\right)$. We use the expression for the effective magnetic field combined with the requirement for the eigenvectors of $H_{\textrm{eff}}$ to derive two implicit equations for the magnetic- and electric field in the $\hat{x}'$-direction as a function of the local magnetic field $\delta B$ in the $\hat{z}$-direction:
\bea
n_{x'}-n_z\frac{\gamma \alpha_1 E_{x'}}{\lambda_+} + n_z^2\frac{\gamma\beta_1 E_{x'}}{\lambda_-}
& = & \left(1+\frac{\delta B}{B_0} \right)\frac{-2c}{1+c^2}\\
n_z+n_{x'}\frac{\gamma \alpha_1 E_{x'}}{\lambda_+} - n_{x'} n_z\frac{\gamma\beta_1 E_{x'}}{\lambda_-}
 & = & \left(1+\frac{\delta B}{B_0} \right)\frac{1-c^2}{1+c^2},
\eea
where we have defined $\gamma \equiv 2e\hbar /E_Z$ and $c \equiv \frac{eE_{x'}}{m\omega_0^2}\left( \frac{1}{\lambda_+}+\frac{n_z\alpha_2}{\lambda_-}-(n_{x'}^2+n_z^2)\frac{\beta_2^{'}}{\lambda_+}\right)\sigma_{y'}$.

\begin{figure}
\centering
\includegraphics[width=0.98\columnwidth]{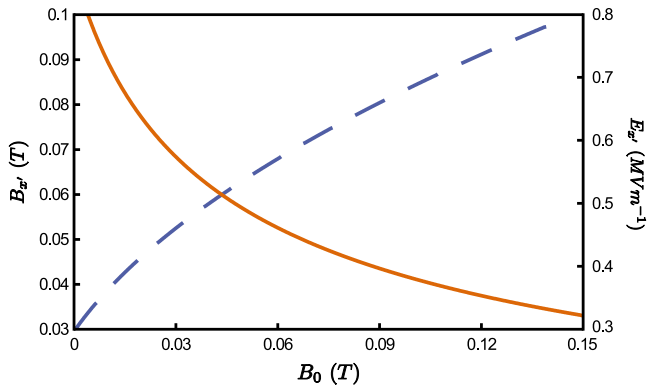}
\caption[]{ $E_{x'}$ (solid line) and $B_{x'}$ (dotted line) vs. $B_0$ for a local magnetic field $\delta B$ corresponding to a Zeeman splitting of $1$ $\mu eV$. Parameters used are 
$\hbar \omega_0 = 1$ meV~\cite{hans07}, $m^{*}=0.067 m_e$
(for GaAs~\cite{hans07}), $\lam_+ = 2\, \mu$m, $\lam_- = 10\,  
\mu$m~\cite{mill03,nowa07}, and $n_z=1$.}
\label{fig:results}
\end{figure}

Fig.~\ref{fig:results} shows the required fields $B_{x'}$ and $E_{x'}$ as a function of $B_0$ for typical experimental parameters and a local Zeeman splitting of $1$ $\mu eV$ ~\cite{parfields}.
We see that in order to have an additional Zeeman splitting of 1 $\mu$eV (as assumed in
Fig.~\ref{fig:survival}), e.g. at $B_0$ = 0.05T magnetic and electric fields $B_{x'} \sim 63$ mT and 
$E_{x'} \sim 4.6\, 10^{5}$ Vm$^{-1}$ are required. Although these values of $B_{x'}$ and $E_{x'}$ 
are (somewhat) larger than those that have
been used so far in experiments~\cite{lair07,nowa07}, they may well become available in the near future. 
Alternatively, one could use larger $B_0$ (corresponding to larger $B_{x'}$ 
and smaller $E_{x'}$) or smaller $\delta B_z$ (corresponding to smaller $B_{x'}$ 
and $E_{x'}$, but also smaller success probability). In addition, we note that in order to preserve 
the qubit's quantum state, the fields need to be switched on and off adiabatically.
The corresponding switching time $T$ of e.g. the tunable magnetic field then has to fulfill~\cite{mess99}
$ T \gg \frac{\hbar n_{x'}}{4 n_z |g^{*}| \mu_B B_0} \sim 0.1$ ns. 
This timescale is well within experimental reach and compatible with the operation time (duration
of electron transfer) $\approx 3.5$ ns that we estimated above.

Another method to generate different Zeeman fields in neighboring dots is by using 
a slanting magnetic field. The latter has recently been demonstrated for the first time by 
integrating a microsize ferromagnet in a double dot device~\cite{pior08}.
As a result of the slanting field, the orbital and spin degrees of freedom become hybridized,
leading to an effective mixed charge-spin two-level system, where the role of the 
coin is played by the pseudospin instead of the real spin. 
A global magnetic field $\sim 2$ T magnetizes the ferromagnet and its inhomogeneity 
leads to a different Zeeman field in the two quantum dots of $\delta B_z \sim 10$ mT 
($|\Del_z - \Del_z^{\prime}| \approx  4 \cdot 10^{-26}$ J~\cite{pior08}). 
An advantage of this method compared to using 
spin-orbit interaction (as discussed above) is that no 
tunable magnetic fields but only tunable gate voltages are needed. A disadvantage is that
the qubit is likely to be more sensitive to orbital decoherence. 

Finally, we briefly discuss the question of how to extend our model from 3 quantum dots to a longer 
one-dimensional chain.
In order for the electron to perform a quantum walk along the chain, opening and closing of tunnel 
barriers and aligning energy levels has to be applied at every position where the particle 
has a finite probability of being found. In addition, not only a step operator, but also a 
reinitialization operator 
$C$ of the coin (spin) degree of freedom is needed~\cite{ahar93}. 
A common choice for $C$ is the Hadamard operator~\cite{kemp03}, which can be implemented
by two coherent rotations around different axes~\cite{kopp06,nowa07,impl}. The dynamics of a quantum
walk of electrons along a one-dimensional chain of quantum dots in the presence of decoherence 
remains an interesting question for future research.

In conclusion, we have proposed an implementation of a discrete quantum walk step
operator in a solid-state nanostructure consisting of a linear chain of 3 quantum dots. 
For currently available
techniques to locally create enhanced Zeeman splitting and taking into account decoherence 
due to quantum charge fluctuations, we have analyzed the probability for coherent transfer 
of the electron to the left or right (conditioned on its spin), and predict that
$>$ 90 \% success probability is feasible. We hope that our results will stimulate a 
proof-of-principle experimental demonstration of the discrete quantum walk step operator 
in a solid-state nanosystem.

This work has been supported by the Netherlands Organisation for
Scientific Research (NWO).


\begin{thebibliography}{99}
\bibitem{dephasing}
In Ref.~\cite{tayl07} the effects of dephasing are taken into account in 
a phenomenological way and modeled by a dephasing rate $\tau_0$.
\bibitem{limitsum}
The sum extends over terms with $\omega_{ab} - \omega_{cd} \ll 1/\Del t$
(secular constraint) where $\Del t$ is the timescale of the Markovian course-grained 
evolution~\cite{cohe92}.
\bibitem{parfields}
Similar values for $B_{x'}$ and $E_{x'}$ are found for an in-plane global magnetic field and out-of-plane local fields.
\end{thebibliography}
\end{document}